# Magnetic Proximity Effect in an Antiferromagnetic Insulator/Topological Insulator Heterostructure with Sharp Interface

Yuxin Liu (刘宇新)[1,2], Xuefan Niu (牛雪翻)[1,2], Rencong Zhang (张仁聪)[1,2], Qinghua Zhang (张庆华)[1,2], Jing Teng (滕静)[1,2,**], and Yongqing Li (李永庆)[1,2,**]

[1]*Beijing National Laboratory for Condensed Matter Physics, Institute of Physics, Chinese Academy of Sciences, Beijing 100190*

[2]*School of Physical Sciences, University of Chinese Academy of Sciences, Beijing 100190*

%Phone: 15611530950; 13126585008

**Correspondence authors: Jing Teng (滕静), Yongqing Li (李永庆)

Email: jteng@iphy.ac.cn; yqli@iphy.ac.cn

**Abstract**

We report an experimental study of electron transport properties of MnSe/$(Bi,Sb)_2Te_3$ heterostructures, in which MnSe is an antiferromagnetic insulator, and $(Bi,Sb)_2Te_3$ is a three-dimensional topological insulator (TI). Strong magnetic proximity effect is manifested in the measurements of the Hall effect and longitudinal resistances. Our analysis shows that the gate voltage can substantially modify the anomalous Hall conductance, which exceeds 0.1 $e^2/h$ at temperature $T$ = 1.6 K and magnetic field $\mu_0H$ = 5 T, despite that only the top TI surface is in proximity to MnSe. This work suggests that heterostructures based on antiferromagnetic insulators provide a promising platform for investigating a wide range of topological spintronic phenomena.

Breaking the time-reversal symmetry (TRS) in topological insulators (TIs) can produce a lot of fascinating quantum phenomena, such as the quantum anomalous Hall effect (QAHE)[1-4], topological magnetoelectric effect[3], and chiral Majorana edge modes[5]. The most convenient method for breaking the TRS is magnetic doping, that has been utilized to demonstrate the QAHE for the first time[2]. However, magnetic doping introduces considerable disorder into the TI surface states due to random distribution of the magnetic dopants, creating a serious obstacle to pursuing novel quantum physics at low energy scale. An alternative approach that can overcome this problem is to build TI-based magnetic heterostructures, in which a TI surface is brought in proximity to a ferromagnetic or ferrimagnetic insulator with perpendicular anisotropy[3]. To date, many magnetic insulators (MIs), including EuS[6], $Y_3Fe_5O_{12}$[7, 8], $Tm_3Fe_5O_{12}$[9], $BaFe_{12}O_{19}$[10], $Cr_2Ge_2Te_6$[11] and (Zn,Cr)Te[12], have been exploited to fabricate TI/MI heterostructures. However, interfacial magnetic interactions in most of these structures are still too weak to allow for observation of an anomalous Hall conductance close to $1/2\ e^2/h$, a hallmark of a sizable mass gap opened in the otherwise gapless Dirac surface states, and also a prerequisite for many interesting experimental proposals[1, 3]. The only exception is $(Zn,Cr)Te/(Bi,Sb)_2Te_3/(Zn,Cr)Te$ sandwich structure, in which the quantized anomalous Hall effect has recently been observed at 30 mK[12]. This structure is, however, also susceptible to the disorder effect, because Cr-doping is utilized to introduce magnetic order to the magnetic insulator layer.

Another approach to breaking the TRS in TI surface states is to construct heterostructures with an antiferromagnetic insulator. In comparison to ferromagnetic insulators, antiferromagnetic insulators possess several advantages, such as absence of magnetic stray fields, robustness against external perturbations, ultrahigh speed in magnetic moment switching, and abundance in material types[13]. Based on the first principles calculations, Luo *et al.*[14] and Eremeev *et al.*[15] independently predicted that in MI/TI heterostructures based on MnSe, a type II antiferromagnetic insulator with a Neel temperature of 135 K[16], the magnetic proximity effect can open an energy gap up to ~ 54 meV in the TI surface states. Subsequently, Matetskiy *et al.* observed a gap of ~ 90 meV in a $MnSe/Bi_2Se_3$ heterostructure with an angular resolved photoemission spectroscopy[17]. Recently, some of us prepared $MnSe/(Bi_{1-x}Sb_x)_2Te_3$ heterostructures, and found that near the TI/MI interface, $Mn(Bi,Sb)_2(Te,Se)_4$ septuple layers formed due to intercalation of MnSe layers into the $(Bi,Sb)_2Te_3$ quintuple layers[18]. However, the anomalous Hall effect in the so-called magnetically extended heterostructures was found to be quite weak, with the maximum $\sigma_{AH}$ less than 0.01 $e^2/h$ [18].

In this work, we have grown MnSe/(Bi,Sb)$_2$Te$_3$ heterostructures with an atomically sharp interface by molecular beam epitaxy (MBE), and the anomalous Hall conductance greater than 0.1 $e^2/h$ have been realized. The strong magnetic proximity effect is supported by the gate voltage dependences of the anomalous Hall effect, longitudinal resistance, and magnetoresistance.

The MnSe/(Bi$_{1-x}$Sb$_x$)$_2$Te$_3$ heterostructures were grown on 0.3 mm thick heat-treated SrTiO$_3$ (111) substrates in a MBE system with a base pressure of $3 \times 10^{-10}$ mbar. The Sb/Bi composition ratio was calibrated with quartz crystal microbalance and set at $x$ = 0.55 throughout this work. The (Bi,Sb)$_2$Te$_3$ layer, usually less than 10 nm thick, was first grown at 270°C, followed by growth of the MnSe layer at 300°C. Reflection-high-energy electron diffraction (RHEED) was used to monitor the epitaxial process. When the deposition process was switched from (Bi,Sb)$_2$Te$_3$ to MnSe, a rapid change in the RHEED pattern could be observed. As shown in Fig. 1(a), the streak spacing of MnSe is about 10% wider than that of the (Bi,Sb)$_2$Te$_3$ layer, consistent with the lattice constants in the literature [MnSe(111): 3.86 Å[19], and (Bi$_{0.5}$Sb$_{0.5}$)$_2$Te$_3$: 4.33 Å[20]]. Fig. 1(b) displays a transmission electron microscopy cross-sectional image, in which the (Bi,Sb)$_2$Te$_3$ and MnSe layers are separated by an atomically sharp interface. This is in stark contrast to the previously reported MnSe/(Bi,Sb)$_2$Te$_3$ heterostructures, in which one or two Mn(Bi,Sb)$_2$(Te,Se)$_4$ septuple layers formed by intercalation of MnSe into the (Bi,Sb)$_2$Te$_3$ quintuple layers, were found to exist near the interface, presumably due to a much higher substrate temperature (~370 °C) during the growth[18].

In the following, we focus on the electron transport data acquired from a MnSe/(Bi,Sb)$_2$Te$_3$ (3 nm/8 nm thick) heterostructure (sample A). Other samples with similar structures have also been measured, and the results are consistent with those presented below. These samples were patterned into Hall bars of millimeter size by hand with a wooden toothpick in order to avoid the sample deterioration during the conventional lithographic process. The transport measurements were performed in a $^4$He vapor flow cryostat with a base temperature of 1.6 K. A back-gate voltage was sometimes applied to the back of the SrTiO$_3$ substrate to tune the chemical potential in the TI layer. A magnetic field was applied perpendicular to the heterostructure surface with a superconducting magnet for the results shown below, but measurements at other field orientations were also carried out using an *in-situ* sample rotator.

Fig. 1(c) depicts the temperature dependence of longitudinal resistance $R_{xx}$ of the heterostructure. As the sample is cooled from room temperature to about 96 K, $R_{xx}$ increases monotonically. This can be attributed to the thermal activation process commonly seen in semiconductors. From 96 K to 20 K, a metallic behavior arises as a consequence of progressive freezing-out of bulk carriers and increasing contribution from the surface states. The resistance upturn at $T < 20$ K is similar to those observed in magnetically doped TIs[11, 21]. The upturn is more pronounced than its counterparts in non-magnetic TIs. The latter can be attributed to the Altshuler-Aronov electron-electron interaction, leading to a $\ln T$-type correction to the longitudinal resistance[22], whereas in the former, the broken TRS suppresses the weak antilocalization (WAL) effect, resulting in stronger $T$-dependence of $R_{xx}$ at liquid helium temperatures.

Fig. 2 shows the magnetotransport properties of the heterostructure at $T$=1.65 K with the gate electrode grounded. The magnetoresistance is positive with a cusp-like dip at zero magnetic field [Fig. 2(a)]. Similar results have often been seen in non-magnetic TIs, and can be ascribed to the WAL effect[22, 23]. The low-field magnetoconductance $\Delta\sigma(B)$, can be described with the simplified Hikami-Larkin-Nagaoka (HLN) equation[24]:

$$\Delta\sigma = \sigma(B) - \sigma(0) = -\alpha \cdot \frac{e^2}{\pi h}\left[\psi\left(\frac{1}{2} + \frac{B_\phi}{B}\right) - \ln\left(\frac{B_\phi}{B}\right)\right], \qquad (1)$$

where $\psi$ is the digamma function, $B_\phi = \hbar/(4De\tau_\phi)$ is a characteristic field relate to electron dephasing time $\tau_\phi$, and $D$ is the diffusion coefficient. Prefactor $\alpha$ is 1/2 for a single channel transport in 2D Dirac electron systems (e.g. a TI surface) [22, 23].

As shown in Fig. 2(b), the magnetoconductance data can be fitted quite well to the HLN equation with prefactor $\alpha = 0.46$. Based on the Hall effect measurements, the charge carriers on the top and bottom surfaces are electrons and holes, respectively, and their densities are estimated to be only a few $10^{12}$ cm$^{-2}$. These two surfaces are therefore expected to contribute separately to the transport. Given the fact that the total magnetoconductance is close to that of a single gapless TI surface, the contribution from the top surface, which interfaces with the MnSe layer, must be overwhelmed by the bottom surface. According to the massive Dirac fermion model, the WAL effect can be suppressed if the magnetic proximity effect opens a sizable gap in the surface states and the Fermi level is not far away from the gap[25].

The Hall resistance data recorded at $T = 1.6$ K is depicted in Fig. 2(a), in which a hysteresis loop is visible for the magnetic fields lower than 2 T. To rule out the possible contribution of Mn-doping, we have performed control experiments on Mn-doped (Bi,Sb)$_2$Te$_3$ thin films, in which the Hall resistances were found to have an opposite sign to the MnSe/(Bi,Sb)$_2$Te$_3$ heterostructures. Fig. 2(c) further shows that the hysteresis starts to disappear at about 10 K. At higher temperatures, the Hall effect curves remain nonlinear. This can be attributed to the parallel transport in the top and bottom surfaces, which in general do not have the same carrier densities or mobilities due to their different interfaces. In addition, the field dependence of the magnetization of MnSe may also be responsible for to the nonlinear Hall effect, since it introduces an anomalous Hall component $R_{AH}$ to the measured Hall resistance $R_{xy}$ via the interface proximity effect. The nonlinearities in the ordinary and anomalous Hall resistances thus pose a daunting challenge to the separation of the two Hall components.

In this work, we use the following method to extract $R_{AH}$. First, the Hall resistances measured in high magnetic fields are fitted to the following nonlinear function:

$$R_{xy}(B) = R_{AH} + R_{OH} = R_{AH} + R_H B + R_3 B^3 + O(B^5), \qquad (2)$$

where the cubic term is the lowest order correction that can take account of the nonlinearity in the ordinary Hall effect. Since the carrier mobilities on the top and bottom surfaces ($\mu_1$, $\mu_2$) are very low for the samples of interest here, $\mu_1 B \ll 1$ and $\mu_2 B \ll 1$ should be satisfied. This justifies dropping out the $B^5$ and higher order terms in Eq. (2) within the framework of two-band semiclassical transport[26]. Subtracting $R_{OH}(B) = R_H B + R_3 B^3$ from the measured $R_{xy}(B)$ for the entire range of magnetic fields, one obtains the $R_{AH}$ data shown in Fig. 2(d), in which the high field fit is limited to a field range of 7-9 T. A hidden assumption in the above procedure is that the magnetization of MnSe is close to saturation at $B > 7$ T. The obtained saturation value of $R_{AH}$ is 471 Ω, corresponding to an anomalous Hall conductance $\sigma_{AH}^s = 0.074\ e^2/h$. We also tried similar analysis with a field range of 5-9 T, which yields $\sigma_{AH}^s = 0.069\ e^2/h$, only 7% smaller than that obtained with the fit of 7-9 T. Nevertheless, it should be noted that the magnetization of MnSe might not reach saturation at 7-9 T, and in this case the actual $\sigma_{AH}^s$ would be even larger than the value obtained above. Such a large saturation field is presumably related to the antiferromagnetic nature of MnSe. Direct magnetization measurement is, however, needed to determine the saturation field of the MnSe thin films. The result of this challenging experiment will be reported elsewhere.

Shown in Fig. 3(a) are the Hall effect curves taken at $T=1.6$ K and different back-gate voltages. Apparently, the dominant charge carriers change from $p$-type to $n$-type, as $V_G$ increases from −100 V to 100 V. After subtracting the ordinary Hall resistance background by performing the procedure described above, the anomalous Hall resistances at various gate voltages can be obtained. The $\sigma_{AH}^s$ values, converted from the high field $R_{AH}$ for each gate voltage, are displayed in Fig. 3(b). A striking feature is the monotonic increase in $\sigma_{AH}^s$ with increasing $V_G$ for the full range of gate voltages, in contrast to the pronounced $R_{xx}$ maximum appearing at $V_G = 1.1$ V. If only the top surface participated in the transport, a $\sigma_{AH}^s$ maximum would coincide with the $R_{xx}$ maximum. The absence of the $\sigma_{AH}^s$ maximum hence suggests that even for the highest gate voltage, the Fermi level on the top surface still resides below the Dirac mass gap induced by the interface proximity effect. In this case, the top and bottom surfaces are of p-type and n-type, respectively. Reducing the gate voltage will further increase the hole density on the top surface. Assuming a monotonic dependence of the surface conductivity on the hole density, one thus expects that the $R_{xx}$ peak at $V_G$= 1.1 V is an indicator for the Fermi level on the bottom surface passing by the Dirac point from above as $V_G$ decreases. Therefore, both the top and bottom surfaces are of p-type at $V_G = -210$ V, consistent with the high hole densities revealed by the Hall resistance data.

Fig. 4(a) schematically illustrates the band diagrams $V_G = 210$ V and $-210$ V based on the above analysis. They are further supported by the magnetotransport data shown in Fig. 4(b) and (c). In case of large positive gate voltages (e.g. $V_G = 210$ V), the Fermi level in the TI layer is located in the bulk band gap, so the top and bottom surfaces contribute independently. Since the WAL in the top surface is suppressed by the magnetic proximity effect, the cusp-shaped magnetoresistances displayed in Fig. 4(b) originate from the nonmagnetic bottom surface. Fig. 4(c) shows that the low field magnetoconductances for $V_G \geq 50$ V can be fitted well to the HLN equation, and all of the obtained $\alpha$ values are close to 1/2. As discussed above for the results shown in Fig. 2, such negative magnetoconductances can be attributed to the WAL effect in the bottom surface. In contrast, the magnetoresistance curves turn into "V" shape for sufficiently large negative gate voltage. As depicted in Fig. 4(b), the corresponding low field magnetoconductances are parabola-shaped for $V_G \leq -50$ V , in stark contrast to the cusp shape observed at large positive gate voltages. This can be explained by the fact that the Dirac point in $(Bi,Sb)_2Te_3$ is located very close to the top of valence band[27]. A sheet hole density of a few $10^{12}$ cm$^{-2}$ is sufficient to lower the chemical potential into the bulk valence band. The p-type bulk carriers can couple to both the top and bottom surfaces, and hence extend the magnetic proximity effect at the top interface across the entire TI layer, resulting in the suppression of WAL in the bottom surface states. It is also noteworthy that the magnetoresistances at $V_G \leq -50$ V are linear for a wide range of magnetic fields. As shown in Fig. 4(b), the linear magnetoresistance is most pronounced at $V_G = -50$ V, at which the bulk states have started to participate in the transport process. Therefore, the linear magnetoresistance observed here cannot be attributed to the quantum linear magnetoresistance of the surface states[28, 29], consistent with previous work on $(Bi,Sb)_2Te_3$ thin films[30].

In summary, we have grown $MnSe/(Bi,Sb)_2Te_3$ heterostructures with atomic sharp interfaces and studied their transport properties systematically. Pronounced anomalous Hall effect has been observed, and it is shown that the anomalous Hall conductivity becomes larger with increasing gate voltage, and can reach as high as 0.12 $e^2/h$ at $V_G = 210$ V, even though the longitudinal resistance exhibits a maximum near the zero gate-voltage. The Hall effect measurements also suggest substantial charge transfer at the $MnSe/(Bi,Sb)_2Te_3$ interface, making the top TI surface p-type for all gate voltages. Such an analysis is supported by the crossover of the magnetoresistance curve from the cusp shape due to the single channel WAL effect to the "V" shape featuring parabolic field dependence in low magnetic fields and linear dependence in higher fields. Our work also suggests that further optimization of the growth condition may allow for observation of the QAHE in MnSe/BST/MnSe sandwich structures.

Corresponding authors: *jteng@iphy.ac.cn, †yqli@iphy.ac.cn

Acknowledgements: This work was supported by National Key Research and Development Program (Project No. 2016YFA0300600), National Science Foundation of China (Project No. 11961141011), and the Strategic Priority Research Program of Chinese Academy of Sciences (Project No.XDB28000000).

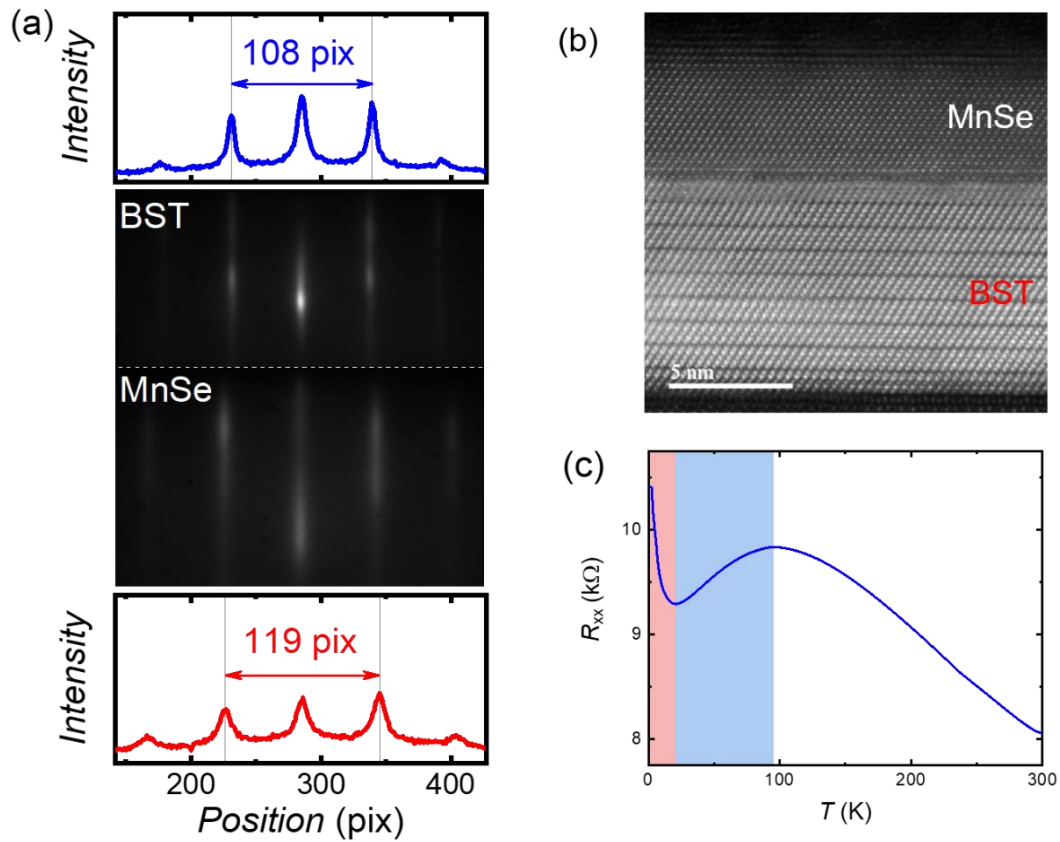

Fig. 1. Basic characterization of MnSe/(Bi,Sb)$_2$Te$_3$ heterostructures grown on STO(111) substrates. (a) RHEED patterns of (Bi,Sb)$_2$Te$_3$ (BST) and MnSe layers and the corresponding intensity peaks. (b) Cross-sectional TEM image of a MnSe/BST (6

nm/9 nm) heterostructure. (c) Temperature dependence of $R_{xx}$ of a MnSe/BST (3 nm/8 nm) heterostructure.

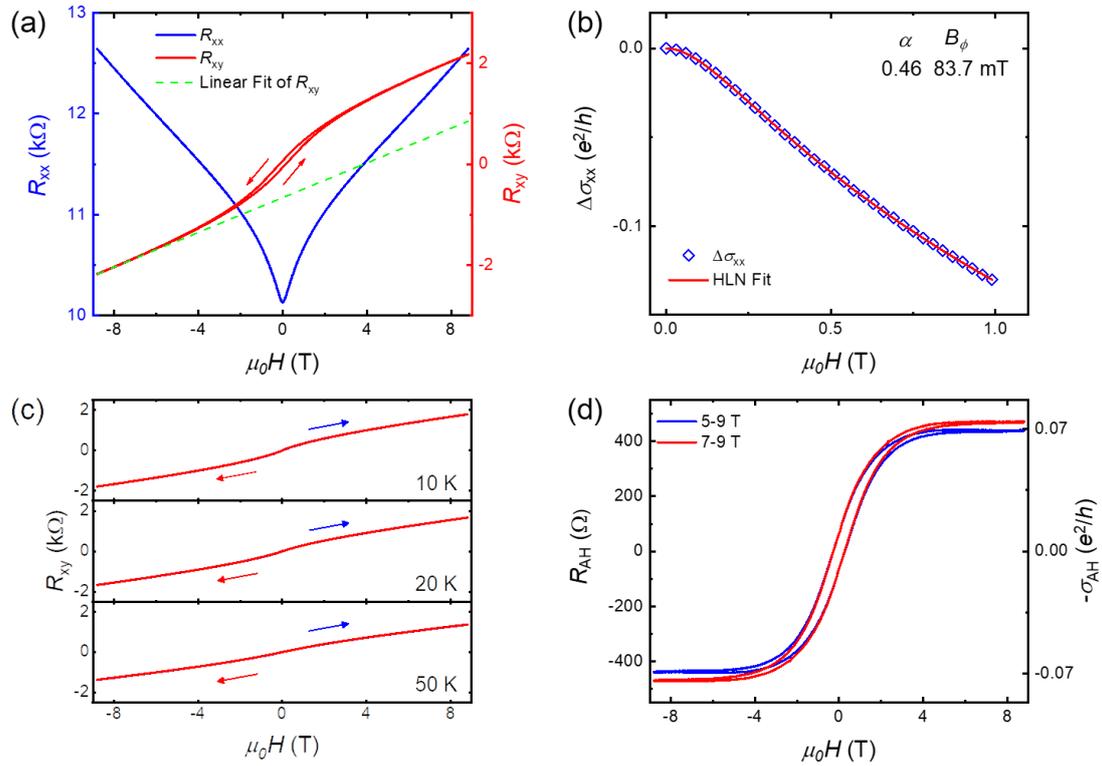

Fig. 2. Transport properties of the MnSe/BST heterostructure (sample A) without gating. (a) Magnetic field dependence of $R_{xx}$ (blue) and $R_{xy}$ (red) at $T$ = 1.6 K. The arrows (red) denote magnetic field sweeping direction. The dashed line (green) is obtained from a linear fit of the $R_{xy}$ data from −8.8 T to −8 T. (b) Magnetoconductivity curve (symbols) and its fit to the HLN equation (line). (c) Hall effect curves taken at 10 K, 20 K and 50 K. (d) Anomalous Hall resistances extracted by subtracting the nonlinear background with two different fitting ranges: 5-9 T(blue) and 7-9 T(red).

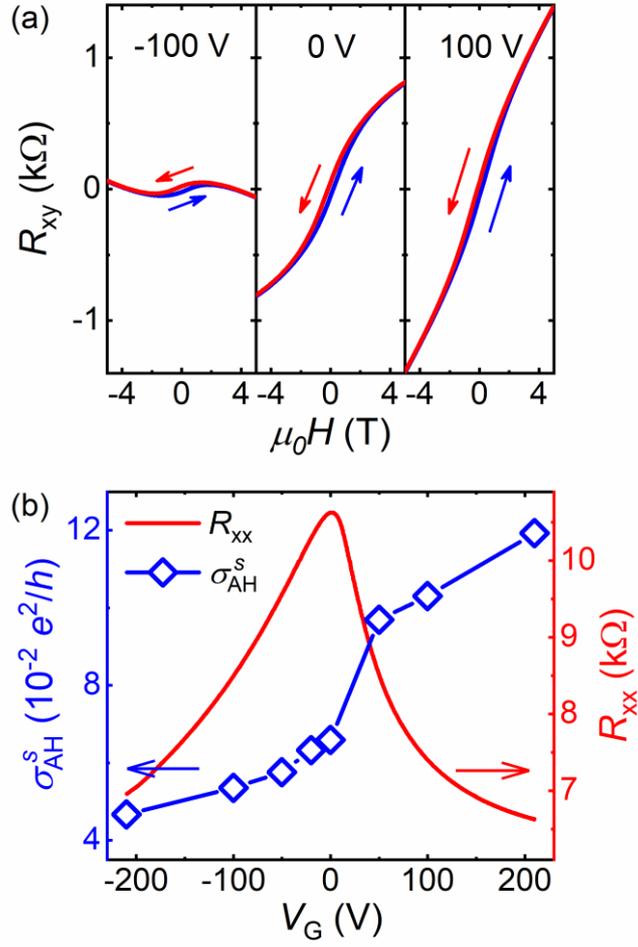

Fig. 3. Gate voltage of the Hall effect in the MnSe/BST heterostructure at $T = 1.6$ K. (a) Hall resistance curves measured at $V_G = -100$ V, 0 V and 100 V. (b) Gate-voltage dependence of $R_{xx}$ at zero magnetic field (red line) and the anomalous Hall conductivity $\sigma_{AH}^s$ at $\mu_0 H = 8.8$ T (blue diamonds).

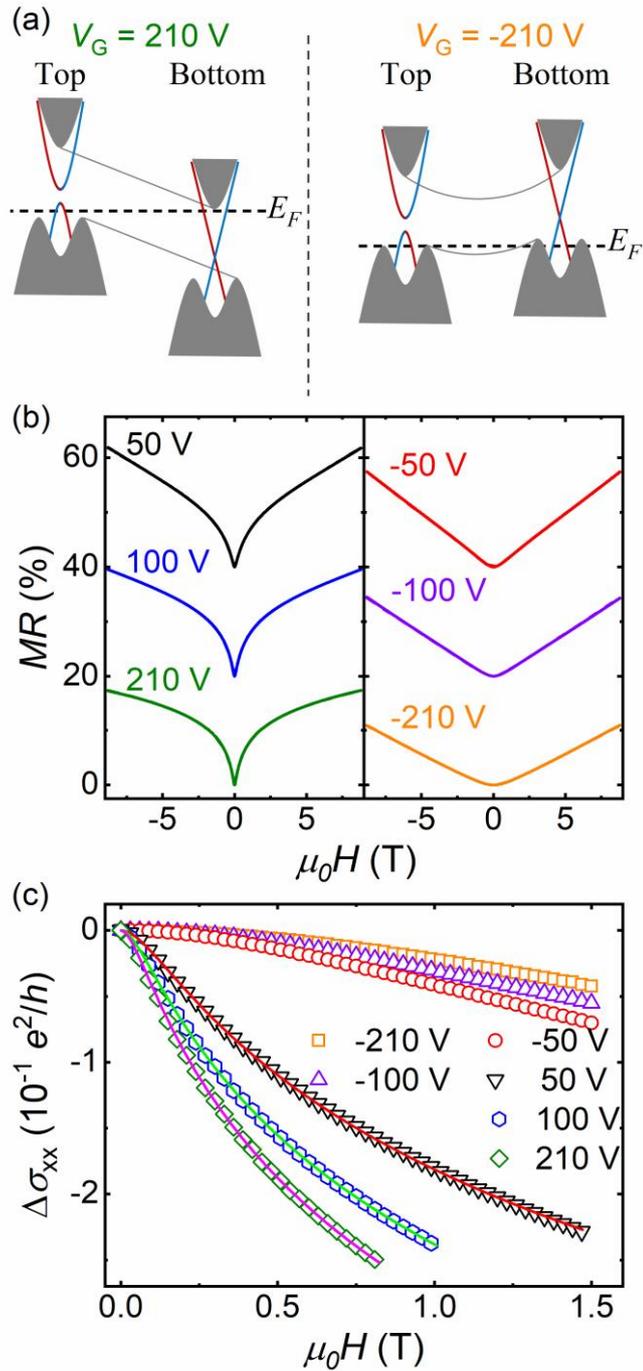

Fig. 4. Band diagrams and the gate-voltage dependence of longitudinal transport properties of the MnSe/BST heterostructure. (a) Schematic band diagrams of the BST layer for $V_G = 210$ V and $-210$ V. (b) Magnetoresistances recorded at six different gate voltages. Some of the MR curves are shifted vertically for clarity. (c) Corresponding magnetoconductance curves (symbols) taken at the same voltages.